\documentclass[aps,prc,twocolumn,showpacs,superscriptaddress,floatfix,10pt]{revtex4-1}
\usepackage{graphicx}% Include figure files
\usepackage{dcolumn}% Align table columns on decimal point
\usepackage{bm}% bold math
\usepackage{amsfonts}

\begin{document}

\title{Dynamic microscopic study of pre-equilibrium giant resonance excitation\\
and fusion in the reactions $^{132}$Sn+$^{48}$Ca and $^{124}$Sn+$^{40}$Ca}
% Force line breaks with \\

\author{V.E. Oberacker}
\author{A.S. Umar}
\affiliation{Department of Physics and Astronomy, Vanderbilt University, Nashville, Tennessee 37235, USA}
\author{J.A. Maruhn}
\affiliation{Institut f\"ur Theoretische Physik, Goethe-Universit\"at, D-60438 Frankfurt am Main, Germany}
\author{P.-G. Reinhard}
\affiliation{Institut f\"ur Theoretische Physik, Universit\"at Erlangen, D-91054 Erlangen, Germany}

\date{\today}

%------------------------------------------------------------------------------

\begin{abstract}
We study pre-equilibrium giant dipole resonance excitation and fusion 
in the neutron-rich system $^{132}$Sn+$^{48}$Ca at energies near the
Coulomb barrier, and we compare photon yields and total fusion cross sections to those
of the stable system $^{124}$Sn+$^{40}$Ca. The dynamic microscopic calculations
are carried out on a three-dimensional lattice using both the Time-Dependent
Hartree-Fock method and the Density Constrained TDHF
method. We demonstrate that the peak of the GDR excitation spectrum
occurs at a substantially lower energy than expected for an equilibrated system,
thus reflecting the very large prolate elongation of the dinuclear
complex during the early stages of fusion. Our theoretical fusion cross-sections for both
systems agree reasonably well with recent data measured at HRIBF.
\end{abstract}
% PACS, the Physics and Astronomy Classification Scheme.
\maketitle

% ------------------------------------------------------------------------------

\section{Introduction}

Heavy-ion reactions at radioactive ion beam (RIB) facilities enable us to synthesize
new exotic neutron-rich nuclei far away from the valley of stability and to study
their physical properties. At the HRIBF facility a series of experiments has been carried out
with radioactive $^{132}$Sn beams on various targets~\cite{Li03,Li08,Li11a,KR12}.

One of the major open questions in fusion reactions of exotic neutron-rich nuclei
is the dependence of the fusion cross section (and related observables) on the neutron excess,
or equivalently on the total isospin quantum number $T_z = (Z-N)/2$. 
To reveal possible systematic trends requires both theoretical
and experimental studies with a wide variety of projectile and target
combinations which are expected to become available at current and future 
RIB facilities. To be able to pin down the isospin dependence in a 
fully microscopic theory, it is desirable to choose collision partners
which are as simple as possible: projectile and target nuclei should be
spherical in their ground state, and the compound nucleus should have a high fission barrier
so that the fission component can be ignored (at least at lower beam energies). Ideal
reaction partners satisfying these criteria include the neutron-rich system $^{132}$Sn+$^{48}$Ca
and the stable system $^{124}$Sn+$^{40}$Ca which we investigate in this paper.

If the reaction partners have appreciably different $N/Z$ ratios, the proton and
neutron centers of mass of the reacting system may not coincide during the early stages of
the collision.
In this case, collective oscillations of protons against
neutrons on the way to fusion might occur, the so-called
pre-equilibrium giant dipole resonance (GDR)~\cite{CT93,SF96,BB01,SCh01,US85}.
In fusion reactions, the shape of the pre-equilibrium di-nuclear complex
exhibits a very large prolate deformation as compared to the shape of
the equilibrated compound nucleus (see e.g. Ref.~\cite{OU10b}, Figures 1 and 2).
Consequently, the $\gamma$-rays emitted during the
early phase of fusion contain information about the shape of the system
and provide a cooling effect along the fusion path as well as insight
into the charge equilibration.

The time-dependent Hartree-Fock (TDHF) theory provides a useful foundation for a
fully microscopic many-body theory of large amplitude collective
motion~\cite{Ne82,Cus85a} including nuclear molecular resonances, deep-inelastic
scattering and heavy-ion fusion, and collective excitations.
The earliest TDHF calculations carried out in the 1970's and 1980's were
severely limited by computer memory and speed. Hence, these calculations
had to make restrictive assumptions such as axial symmetry and the omission of
spin-orbit coupling. During the past several years, with a steady increase in
computational power, it has become feasible to perform TDHF calculations on a
three-dimensional (3D) Cartesian grid with no symmetry restrictions whatsoever
and with much more accurate numerical methods~\cite{UO06,GM08,DD-TDHF,KS10}.
At the same time the quality of effective interactions has also been substantially
improved~\cite{CB98,BH03,Klu09a,HFB14,KL10}. Our TDHF code utilizes the full Skyrme
interaction, including all of the time-odd terms in the mean-field Hamiltonian~\cite{EB75,UO06}.

During the past several years, we have developed the density-constrained TDHF
method (DC-TDHF) for calculating heavy-ion potentials~\cite{UO06a}, and we have applied this method
to calculate fusion and capture cross sections above and below the barrier. So far, we have studied
the systems $^{132}$Sn+$^{64}$Ni~\cite{UO07a}, $^{64}$Ni+$^{64}$Ni~\cite{UO08a},
$^{16}$O+$^{208}$Pb~\cite{UO09b}, $^{132,124}$Sn+$^{96}$Zr~\cite{OU10b},
and we have studied the entrance channel dynamics of hot and cold fusion reactions leading
to superheavy element $Z=112$~\cite{UO10a}. In all cases, we have found
good agreement between the measured fusion cross sections and the DC-TDHF results.
This is rather remarkable given the fact that the only input in DC-TDHF is the 
Skyrme effective N-N interaction, and there are no adjustable parameters.
 
This paper is organized as follows: in Section~II we summarize the theoretical formalism
for pre-equilibrium GDR excitation in heavy-ion fusion reactions
with the TDHF method, and we compare the resulting dipole radiation yields for the
neutron-rich system $^{132}$Sn+$^{48}$Ca to the photon spectra of the stable
system $^{124}$Sn+$^{40}$Ca. 
In Section~III, we study total fusion cross sections for both systems. First, we
carry out unrestricted TDHF runs at energies above the barrier (no tunneling).
Secondly, we calculate the heavy-ion interaction potential $V(R)$ and the
coordinate-dependent mass parameter $M(R)$ utilizing the DC-TDHF method.
By numerical integration of the Schr\"odinger equation for the relative
coordinate $R$ with the {\it Incoming Wave Boundary Condition} (IWBC) method,
we calculate total fusion cross sections both below and above
the barrier, and we compare our theoretical results to recent data measured at HRIBF.
Our conclusions are presented in Section~IV.

% ------------------------------------------------------------------------------

\section{TDHF study of pre-equilibrium GDR excitation}

In the present TDHF calculations we use the Skyrme SLy4 interaction~\cite{CB98} for the nucleons
including all of the time-odd terms in the mean-field Hamiltonian~\cite{UO06}.
The numerical calculations are carried out on a 3D Cartesian lattice. For the two
reactions studied here, the lattice spans $50$~fm along the collision axis and $30-42$~fm in
the other two directions, depending on the impact parameter. Derivative operators on
the lattice are represented by the Basis-Spline collocation method. One of the major
advantages of this method is that we
may use a relatively large grid spacing of $1.0$~fm and nevertheless achieve high numerical
accuracy. First we generate very accurate static HF wave functions for the two nuclei on the
3D grid. The static HF equations are solved with the damped gradient iteration method.
The initial separation of the two nuclei is $22$~fm for central collisions. In the second
step, we apply a boost operator to the single-particle wave functions. The time-propagation
is carried out using a Taylor series expansion (up to orders $10-12$) of the unitary mean-field propagator,
with a time step $\Delta t = 0.4$~fm/c. 

In TDHF, the dissipation of the relative kinetic energy into internal excitations is
due to the collisions of the nucleons with the ``walls'' of the
self-consistent mean-field potential~\cite{Gro75a,Blo78a}. Because the randomization of the
single-particle motion occurs through repeated exchange of
nucleons from one nucleus into the other, the equilibration of
excitations is rather slow, and it is sensitive to the details of the
shape evolution of the composite system. This is in contrast to
most classical pictures of nuclear fusion which generally assume near
instantaneous, isotropic equilibration.

\begin{figure}[!htb]
\includegraphics*[width=8.6cm]{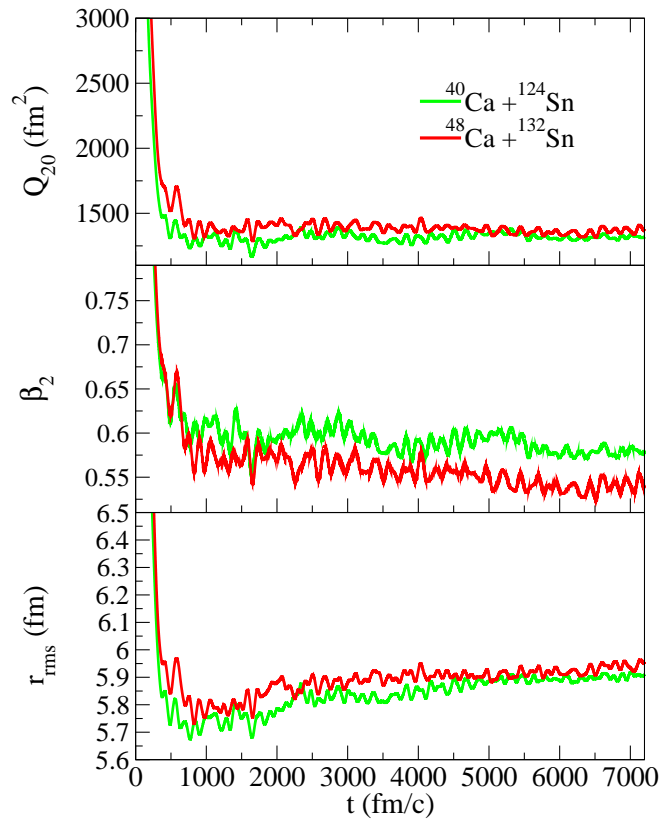}
\caption{\label{fig:fig1} (Color online) Time evolution of isoscalar multipole moments for a head-on collision of
Sn on Ca at $E_\mathrm{cm}=130$~MeV. Two isotopic combinations are considered as indicated.}
\end{figure}

In Fig.~\ref{fig:fig1} we show the time evolution of isoscalar multipole
moments for a head-on collision of Sn on Ca at $E_\mathrm{cm}=130$~MeV. Compared
are the two isotopic combinations $^{132}$Sn+$^{48}$Ca and $^{124}$Sn+$^{40}$Ca.
The isoscalar quadrupole moments $Q_{20}$ and the associated
quadrupole deformations
  \begin{equation}
    \beta_{20}
    =
    \frac{4\pi}{5}\frac{\langle r^2Y_{20}\rangle}{Ar_\mathrm{rms}}
  \end{equation}
show a remarkable stability in the pre-compound state, i.e. 
after $1000$~fm/c. However, this extraordinarily long life-time
may be partially due to the frozen occupations (no self-consistent pairing) in the TDHF 
calculations and partially due to other missing correlations.
The center-of-mass coordinate $x_\mathrm{c.m.}$
moves less than $0.01$ fm.
The isoscalar r.m.s. radius shows clearly a small, but steady
growth. This is most probably caused by nucleon emission in
connection with reflecting boundary conditions. Nucleons which
leave the compound system are spread over all box volume and
deliver unnaturally larger contributions to the radius.
A computation of the radii in an analyzing volume confined
to the vicinity of the compound system would probably show
the opposite, slightly shrinking radii.
\begin{figure}[!htb]
\includegraphics*[width=8.6cm]{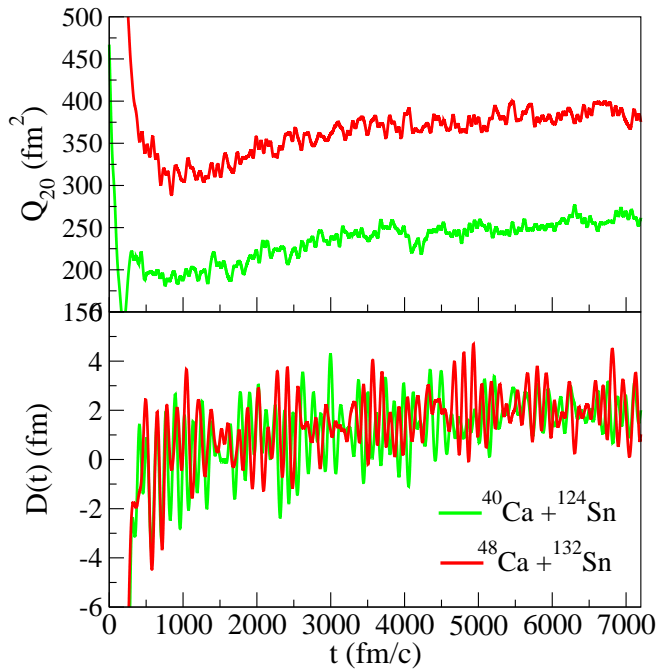}
\caption{\label{fig:fig2} (Color online) Time evolution of isovector multipole moments for a head-on collision of
Sn on Ca at $E_\mathrm{cm}=130$~MeV. Two isotopic combinations are considered as indicated.}
\end{figure}

The excitation of the pre-compound giant dipole mode, which is likely when
the ions have significantly different $N/Z$ ratio, is a reflection of dynamical
equilibration. Dipole excitation results in the emission of $\gamma$ radiation and 
possibly pre-equilibrium particle emission
which in turn provides an important cooling mechanism during the
early stages of the fusion reaction. Several transport simulations like
semiclassical Boltzmann-Nordheim-Vlasov~\cite{CT93,B96} as well as 
microscopic calculations~\cite{SCh07,UO07a,BR09,US85} have confirmed this scenario.

To calculate the electric dipole radiation emitted by the time-dependent
charge distribution, we consider the dipole operator
\begin{equation}
\mathbf{d} = e \sum_{p=1}^Z \mathbf{r}_p \ .
\end{equation}

By adding and subtracting the position vectors of the neutrons we can write this
expression in the form
\begin{equation}
\mathbf{d} = \frac{e}{A} \left [ N \sum_{p=1}^Z \mathbf{r}_p - Z \sum_{n=1}^N \mathbf{r}_n +
              Z \left ( \sum_{p=1}^Z \mathbf{r}_p + \sum_{n=1}^N \mathbf{r}_n \right ) \right ] \ .
\end{equation}
We now impose the condition that the center-of-mass is fixed at the origin of the
coordinate system at all times. As a result, the last term in the round brackets
vanishes and we obtain the dipole operator in the form
\begin{equation}
\mathbf{d} = e \frac{NZ}{A} \left [ \frac{1}{Z} \sum_{p=1}^Z \mathbf{r}_p - 
                                    \frac{1}{N} \sum_{n=1}^N \mathbf{r}_n \right ] \ .
\end{equation}
Let us now consider a central collision in $x$-direction and introduce
the quantity
\begin{equation}
D(t)=\frac{NZ}{A} \left [\frac{1}{Z} \sum_{p=1}^Z <x_p(t)> - 
                                    \frac{1}{N} \sum_{n=1}^N <x_n(t)> \right ]
\end{equation}
which represents the expectation value of the $x$-component of the dipole operator
$d_x / e$ taken with the time-dependent TDHF Slater determinant $|\Phi(t)>$.

In Fig.~\ref{fig:fig2} we show the time evolution of the corresponding
isovector quadrupole moment (top panel) and the isovector dipole moment $D(t)$ (bottom panel) for the two systems. 
It is interesting to compare the isovector dipole moments
  $D(t)$ for $^{132}$Sn+$^{48}$Ca and for $^{124}$Sn+$^{40}$Ca.
  The initial amplitude seems to be nearly the same in 
  both cases which may question the argument that the total
  isospin $T_z$ is responsible for the strong dipole excitations.
  We have also computed a collision of the symmetric system
  $^{88}$Sr+$^{88}$Sr which ends in a comparable compound state. 
  There is no isovector dipole excitation in this case as expected.
  There is another interesting feature in that the dipole
  amplitude remains rather stable for $^{132}$Sn+$^{48}$Ca while
  it shows signs of damping for $^{124}$Sn+$^{40}$Ca. 

Following the bremsstrahlung approach developed by Baran et al.~\cite{B96} we
define the dipole acceleration 
\begin{equation}
D''(t) = \frac{d^2 D(t)}{dt^2}
\end{equation}
and introduce its Fourier transform
\begin{equation}
D''(\omega) = \int_{t_{min}}^{t_{max}} D''(t) e^{i \omega t} dt\;.
\end{equation}
Alternatively, for nearly harmonic vibrations one can use the expression $\left|D''(\omega)\right|^2 = \omega^4\left|D(\omega)\right|^2$, with
\begin{equation}
  D(\omega) =
  \int_{t_\mathrm{min}}^{t_\mathrm{max}} D(t) e^{i \omega t}
  \sin^4\left(\pi\frac{t-t_{\mathrm{min}}}{t_{\mathrm{max}}-t_{\mathrm{min}}}\right)dt \;.
\end{equation}
The time filtering $\sin^4$ is used to smooth out peaks coming from finite integration time.
The ``power spectrum'' of the electric dipole radiation is given by~\cite{B96}
\begin{equation}
\label{eq:yield}
\frac{dP}{dE_{\gamma}}=\frac{2 \alpha} {3 \pi E_{\gamma}}
 \left |\frac{1}{c} D''(\omega) \right |^{2} \ ,
\end{equation}
where $\alpha = e^2/(\hbar c) \approx 1/137$ denotes the fine structure constant.

\begin{figure}[!htb]
\includegraphics*[width=6.4cm,angle=-90]{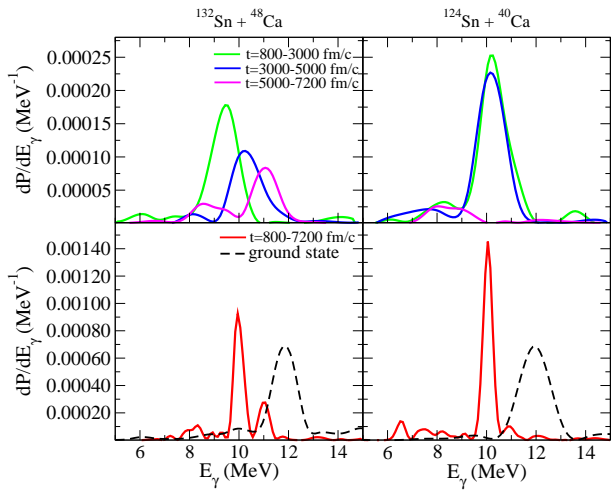}
\caption{\label{fig:d10_E} (Color online)
Power spectrum of pre-equilibrium dipole radiation following
a central collision at $E_\mathrm{c.m.}=130$~MeV of $^{132}$Sn+$^{48}$Ca (left panel)
and  $^{124}$Sn+$^{40}$Ca (right panel).  Results are shown
for spectral analysis over the whole simulation time (lower panels, red)
and over three shorter intervals as indicated (upper panels).
The radiation power spectrum of 
the (deformed) ground state of the compound system
is shown by the dashed curve (lower panels).}
\end{figure}

In Fig.~\ref{fig:d10_E} we show results of TDHF calculations of the photon yield
given by Eq.~\ref{eq:yield}. 
For our calculations the use of $D''(t)$ or the alternate expression with $D(t)$
give essentially the same result.
Compared are the neutron-rich system $^{132}$Sn+$^{48}$Ca
and the stable system $^{124}$Sn+$^{40}$Ca at the same center-of-mass energy.
In performing the full Fourier transform 
we have cut out the approach phase of the two ions corresponding to times $t < t_{min}=800$~fm/c. The
upper limit of the time integration is $t_{max}=7,200$~fm/c which corresponds to $2.16 \times 10^{-20}$ s.
The lower part (red curve) of the Fig.~\ref{fig:d10_E} shows the 
spectrum obtained for the whole time interval $800-7,200$~fm/c.
We observe that the isovector dipole radiation spectrum shows a
peak near $10$~MeV for both isotopic combinations.
Furthermore, we have also divided the total time span into three
intervals and evaluated the power spectra separately for these (see upper panels
in Fig.~\ref{fig:d10_E}. This indicates the time evolution of the power spectra.
The case of
$^{132}$Sn+$^{48}$Ca shows a steady upshift of the mean peak and
decrease of overall strength. The latter is an obvious effect of
damping.
The upshift indicates a decreasing system size. This complies with
the expectation that the true system radii (evaluated in an
analyzing volume) should shrink.
The system $^{124}$Sn+$^{40}$Ca behaves differently. The spectra
stay stable during the first two time intervals and shrink dramatically in the
third time interval which complies with the fact that the dipole amplitude
is visibly smaller at later times (see $D(t)$ panel in Fig.~\ref{fig:fig2}).
In light of the entrance channel $N/Z$ ratios for the two systems, $1.48$ and $1$ for
$^{124}$Sn+$^{40}$Ca stable system and $1.64$ and $1.4$ for the $^{132}$Sn+$^{48}$Ca
neutron-rich system, one can conclude that the system with a larger initial
$N/Z$ differential, namely the stable system, should have the stronger initial pre-equilibrium GDR excitation,
As can be seen from Fig.~\ref{fig:d10_E} this is indeed the case.
On the other hand the presence of extra neutrons for the neutron-rich system seem
to prolong the life of the GDR suggesting a longer equilibration time for this case.

We have also evaluated the dipole radiation spectra originating from
the compound systems in their ground states. Both of these have a quadrupole
deformation
$\beta_{20}\approx 0.3$ which is significantly smaller than the
deformation of the pre-compound state, $\beta_{20}\approx 0.6$.
Figure~\ref{fig:d10_E} (lower panels, dashed black curves) shows also the power
spectra produced by the compound systems. Because of the smaller deformation
the spectra peak at a higher energy, and this peak is 
well separated from the peak of pre-compound stage.

In order to estimate the energy content in the isovector-dipole oscillations,
we have computed the dipole polarizability $\alpha_D$ and find for the pre-compound
system $Z/N = 70/110$ the value $\alpha_D = 30$ fm$^3$. The
corresponding excitation energy is given by $E_{dip} = e^2 {\bar{D}}^2 / (2 \alpha_D)$.
This energy content changes based on what stage of the reaction is considered. During
the initial contact phase, where the dipole excitation may be the dominant form of
collective excitation the amplitude $\bar{D}$ (see Figure~\ref{fig:fig2}) is as large
as $5-6$~fm, which yields $0.6-0.8$~MeV excitation whereas at later stages the
dipole amplitude is in the range of $2-3$~fm resulting in approximately $0.1-0.2$~MeV
excitation.
Our calculations of excitation energy based on the DC-TDHF method (see Ref.~\cite{UOMR09})
for this c.m. energy indicate that the amount of
intrinsic excitation energy starts to rise from zero as the two nuclei make contact
and reach the value of $40$~MeV as the initial composite system is formed.
Thus the fraction of the excitation energy carried by the pre-equilibrium GDR can 
be a large part of the small excitation energy near barrier (by as much as
$40$\% at the barrier peak), while it becomes less than a percent in the interior region.

% ------------------------------------------------------------------------------

\section{Fusion}

\subsection{TDHF calculations at energies above the barrier}

Depending upon beam energy and impact parameter, the dominant reaction channels
are deep-inelastic and fusion reactions~\cite{Arm85aR}. In general, central collisions and
collisions with relatively small impact parameter result in fusion, while 
larger impact parameters give rise to deep-inelastic reactions.
At $E_{\mathrm{c.m.}}$ energies above the potential barrier it is possible
to carry out standard unrestricted TDHF runs (no tunneling). By varying the impact parameter
and plotting the density contours as a function of time, one can easily distinguish
between fusion events (one fragment in the exit channel) and deep-inelastic reactions
(two fragments).

\begin{figure}[!htb]
\includegraphics*[width=8.6cm]{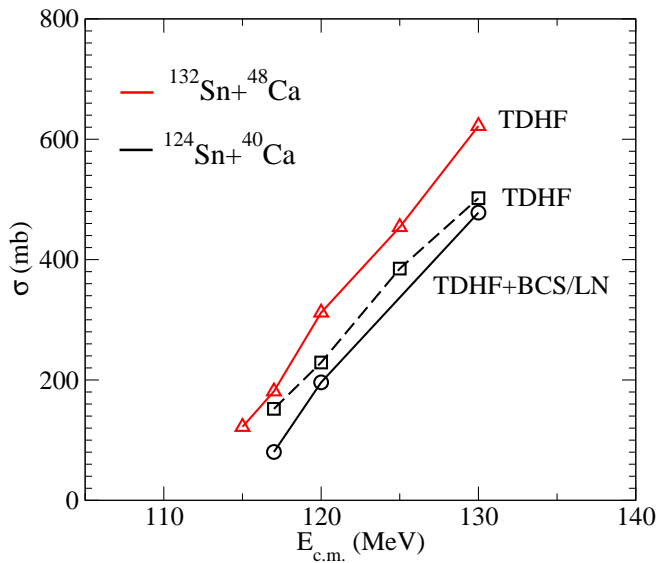}
\caption{\label{fig:fus1} (Color online) TDHF calculation of the total fusion cross
section at energies above the barrier. At fixed $E_\mathrm{c.m.}$ energy, the neutron-rich
system $^{132}$Sn+$^{48}$Ca shows a higher cross section than the stable system $^{124}$Sn+$^{40}$Ca.}
\end{figure}

Let us first consider the case of the neutron-rich system $^{132}$Sn+$^{48}$Ca.
The HF initialization run shows that both nuclei are spherical in their ground state,
in agreement with HFB calculations and with experimental data. At energy
$E_{\mathrm{c.m.}}=130$~MeV we find that impact parameters  $\leq b_{max}=4.45$~fm result
in fusion, while impact parameters $b > 4.45$~fm lead to deep-inelastic reactions.
Using the sharp cut-off model, the fusion cross section is given by
$\sigma_{\mathrm{fus}} = \pi b_{max}^2 = 62.2$~fm$^2 = 622$~mb. This result
is displayed in Fig.~\ref{fig:fus1}. Similar TDHF runs have been carried out at 
energies $E_{\mathrm{c.m.}}=115,117,120,125$~MeV, and the corresponding fusion
cross sections are also shown in the figure.

When we attempted to calculate fusion cross sections for the stable system $^{124}$Sn+$^{40}$Ca
we found that the HF initialization run correctly predicts a spherical ground state for $^{40}$Ca,
but $^{124}$Sn with non-magic neutron number $N=74$ shows a prolate quadrupole deformation
$\beta_2=0.054$ whereas Skyrme-HFB calculations~\cite{DS04,BO05} and experiment show this nucleus to be
spherical. The unphysical shape is due to the lack of pairing in pure HF. Because the
shape of the nucleus is very important for fusion calculations, we have added
BCS/Lipkin-Nogami pairing to the HF initialization of $^{124}$Sn which results in an approximately
spherical shape. During the subsequent TDHF run, we have kept the BCS occupation numbers
frozen. This approximation is necessary, because a full 3D-TDHFB code with
self-consistent pairing for nuclear reactions does not currently exist.

The resulting total fusion cross sections for $^{124}$Sn+$^{40}$Ca are also
shown in Fig.~\ref{fig:fus1}. The solid black line corresponds to TDHF calculations
with added pairing (labeled TDHF+BCS/LN), and the dashed line corresponds to pure
TDHF which results in a larger fusion cross section for the most favorable
orientation of the (unphysical) deformed $^{124}$Sn nucleus.
We notice that if we plot both systems as a function of the
center-of-mass energy, the more massive neutron-rich system shows a larger fusion
cross section. Part of this is a trivial size-effect which can be taking out
by appropriate scaling factors. We will comment on
this issue further when we compare the TDHF cross sections to experimental data
(Fig.~\ref{fig:sigma_scaled} at the end of Section III).

% ---------------------------------------------

\subsection{DC-TDHF heavy-ion interaction potential and mass parameter}

In the absence of a true quantum many-body theory of barrier tunneling, sub-barrier fusion
calculations are reduced to the calculation of a potential barrier between the interacting
nuclei and a subsequent calculation of tunneling through the barrier~\cite{UO07a}.
To date, theoretical studies of fusion cross sections are still dominated
by phenomenological methods such as the coupled-channels (CC) approach~\cite{EJ10,IH09}.
In this approach, one either uses empirical ion-ion potentials
or one calculates the heavy-ion potential with the double-folding method using
measured nuclear densities for projectile and target~\cite{GL79}.
In the latter case, one relies on the ``frozen density'' or ``sudden'' approximation
in which the nuclear densities are unchanged during the computation of the ion-ion
potential as a function of the internuclear distance.
The frozen density approximation ignores dynamical effects such as neck formation
during the nuclear overlap. 
It has been demonstrated that for deep sub-barrier energies the inner part of the
potential barrier plays a very important role~\cite{IH07a}.
 
While phenomenological methods provide a useful starting point for the analysis
of multitudinous fusion data it is desirable to
make contact with the microscopic theories of nuclear structure and reactions.
We have developed a fully microscopic method to extract heavy-ion interaction potentials
$V(R)$ from the TDHF time-evolution of the dinuclear system.
In our DC-TDHF approach~\cite{UO06a}, the time-evolution takes place with no restrictions.
At certain times $t$ during the evolution the instantaneous TDHF density is used to
perform a static Hartree-Fock energy minimization. In a typical DC-TDHF run, we utilize a few
thousand time steps, and the density constraint is applied every $20$ time steps. 
In this density-constrained method
we impose the condition that the total proton and neutron density of the dinuclear system
at the ion-ion separation distance $R(t)$ is equal to the instantaneous TDHF density.
We refer to the minimized energy as the density constrained energy $E_{\mathrm{DC}}(R(t))$.
In the DC-TDHF method the ion-ion interaction potential is given by
\begin{equation}
V(R)=E_{\mathrm{DC}}(R)-E_{\mathrm{A_{1}}}-E_{\mathrm{A_{2}}}\;,
\label{eq:vr}
\end{equation}
where $E_{\mathrm{A_{1}}}$ and $E_{\mathrm{A_{2}}}$ are the binding energies of
the two nuclei obtained with the same effective interaction.
For the calculation of the ion-ion separation distance $R$ we use a hybrid method
as described in Ref.~\cite{UO09b}. At large distances
where a visible neck allows us to identify two fragments we
compute it as the distance between the centers of mass of the ions. For more
compact configurations, we compute $R$ from the mass quadrupole moment $Q_{20}$
as $R=r_0\sqrt{|Q_{20}|}$ where $r_0$ is a scale factor to connect the definition smoothly 
to the large-distance region. 

The interaction potentials calculated with the DC-TDHF method incorporate
all of the dynamical entrance channel effects such as neck formation,
particle exchange, internal excitations, and deformation effects.
While the outer part of the potential barrier is largely determined by
the entrance channel properties, the inner part of the potential barrier
is strongly sensitive to dynamical
effects such as particle transfer and neck formation.

\begin{figure}[!htb]
\includegraphics*[width=8.6cm]{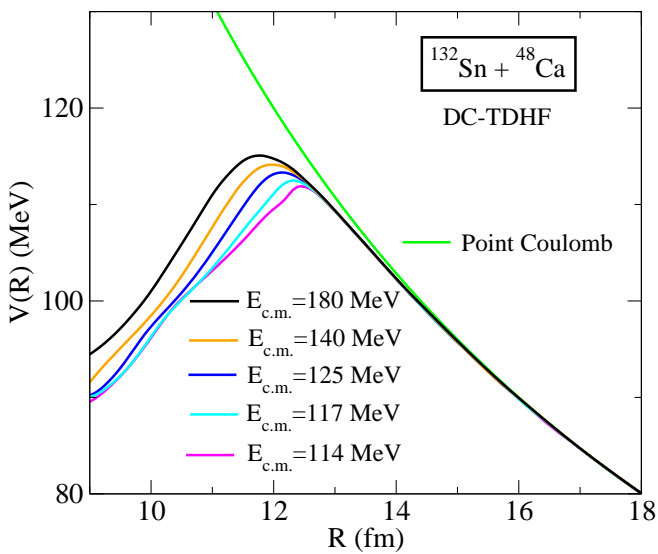}
\caption{\label{fig:pot} (Color online) DC-TDHF calculation of the heavy-ion potential
$V(R)$ for the neutron-rich system $^{132}$Sn+$^{48}$Ca. The energy-dependent potential is 
shown at five $E_\mathrm{c.m.}$ energies. The point Coulomb potential is given for
comparison.}
\end{figure}

In Fig.~\ref{fig:pot} we show the heavy-ion interaction potential $V(R)$
for $^{132}$Sn+$^{48}$Ca calculated with the DC-TDHF method using Eq.~\protect(\ref{eq:vr}).
The heavy-ion potential depends on the center-of-mass energy and is shown at five different
$E_\mathrm{c.m.}$ energies. Our results demonstrate that in these heavy
systems the potential barrier height increases dramatically with increasing
energy $E_\mathrm{c.m.}$, and the barrier peak moves inward towards
smaller $R$-values.
By contrast, DC-TDHF calculations for light ion systems such as $^{16}$O+$^{16}$O
show almost no energy-dependence even if we increase $E_\mathrm{c.m.}$ by a factor
of four~\cite{UOMR09}. Even in reactions between a light and a very heavy nucleus
such as $^{16}$O+$^{208}$Pb, we see only a relatively small energy dependence
of the barrier height and width~\cite{UO09b}.

In addition to the ion-ion potential, it is also possible to compute the corresponding coordinate
dependent mass parameter $M(R)$ using energy conservation at zero impact parameter
\begin{equation}
M(R)=\frac{2[E_{\mathrm{c.m.}}-V(R)]}{\dot{R}^{2}}\;,
\label{eq:mr}
\end{equation}
where the collective velocity $\dot{R}$ is directly obtained from the TDHF evolution and the potential
$V(R)$ from the density constraint calculations.

\begin{figure}[!htb]
\includegraphics*[width=8.6cm]{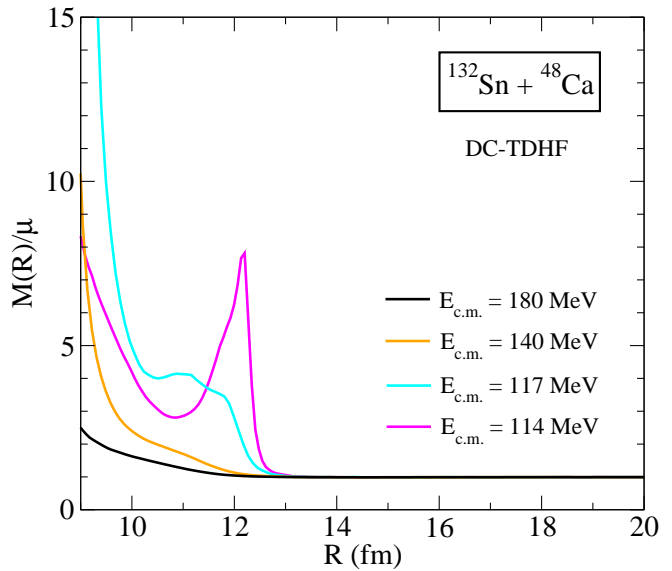}
\caption{\label{fig:mass} (Color online) Microscopic mass parameter $M(R)$ (in
units of the reduced mass $\mu$), calculated at several $E_\mathrm{c.m.}$
energies.}
\end{figure}

In Fig.~\ref{fig:mass} we show the mass parameter $M(R)$ calculated from
Eq.~\protect(\ref{eq:mr}) at four different energies
$E_\mathrm{c.m.}$. As expected, at large distance $R$ the mass $M(R)$ is equal to the
reduced mass $\mu$ of the system. At smaller distances, when the nuclei overlap, the
mass parameter increases in all cases. Like the heavy-ion potential, the mass parameter shows a
strong dependence on energy: at higher energies, $M(R)$ rises smoothly with
decreasing $R$. However, for the lowest energies which are only a few MeV above the
corresponding potential barrier, the mass parameter shows a pronounced peak. This is due
to the fact that the velocity $\dot{R}$ in
the denominator of Eq.~\protect(\ref{eq:mr}) becomes small as the ions approach each other,
reaches its minimum at the top of the barrier, and then increases again at smaller $R$-values.

\begin{figure}[!htb]
\includegraphics*[width=8.6cm]{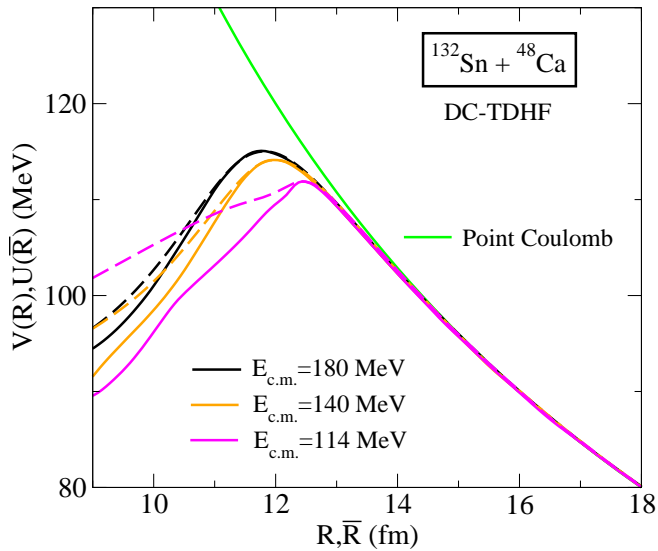}
\caption{\label{fig:pot1} (Color online) Solid lines: original heavy-ion potentials
$V(R)$ corresponding to the mass parameter $M(R)$. Dashed lines: transformed potentials
$U(\bar{R})$ corresponding to the reduced mass $\mu$.}
\end{figure}

The total fusion cross section is given by
\begin{equation}
\sigma_{\mathrm{fus}}(E_{\mathrm{c.m.}}) = \frac{\pi \hbar^2}{2 \mu E_{\mathrm{c.m.}}}
                     \sum_{\ell=0}^{\infty} (2\ell+1) T_{\ell}(E_{\mathrm{c.m.}}) \ .
\label{eq:sigma_fus}
\end{equation}
The potential barrier penetrabilities $T_{\ell}$ are obtained from the numerical
solution of the Schr\"odinger equation for the relative coordinate $R$.
Instead of solving the Schr\"odinger equation with coordinate dependent
mass parameter $M(R)$ for the heavy-ion potential $V(R)$, it is numerically 
advantageous to use the constant reduced mass $\mu$ and to transfer the
coordinate-dependence of the mass to a scaled
potential $U(\bar{R})$ using the scale transformation
\begin{equation}
d\bar{R}=\left(\frac{M(R)}{\mu}\right)^{\frac{1}{2}}dR\;.
\label{eq:mrbar}
\end{equation}
Integration of Eq.~(\ref{eq:mrbar}) yields
\begin{equation}
\bar{R}= f(R) \ \ \ \Longleftrightarrow \ \ \ R=f^{-1}(\bar{R})\;.
\label{eq:rbar}
\end{equation}
As a result of this point transformation, the scaled heavy-ion potential is given by the expression
\begin{equation}
V(R) = V(f^{-1}(\bar{R})) = U(\bar{R}) \;.
\label{eq:Urbar}
\end{equation}
In Fig.~\ref{fig:pot1} we compare the original heavy-ion potential $V(R)$ (solid lines)
corresponding to the mass parameter $M(R)$ to the transformed potential $U(\bar{R})$
(dashed lines) calculated from Eq.~\protect(\ref{eq:Urbar}) which corresponds to the
constant reduced mass $\mu$. We observe that the coordinate-dependent mass changes only 
the interior region of the potential barriers, and this change is most pronounced
at low $E_\mathrm{c.m.}$ energies.

The Schr\"odinger equation corresponding to the constant reduced mass $\mu$ and the scaled
potential $U(\bar{R})$ has the familiar form
\begin{equation}
\left [ \frac{-\hbar^2}{2\mu} \frac{d^2}{d\bar{R}^2} + \frac{\hbar^2 \ell (\ell+1)}{2 \mu \bar{R}^2} + U(\bar{R})
 - E_\mathrm{c.m.} \right] \psi_{\ell}(\bar{R}) = 0 \;.
\label{eq:Schroed1}
\end{equation}
By numerical integration of Eq.~(\ref{eq:Schroed1}) using the well-established
{\it Incoming Wave Boundary Condition} (IWBC) method~\cite{Raw64,HW07}we obtain
the barrier penetrabilities $T_{\ell}$ which determine the total fusion cross
section~(\ref{eq:sigma_fus}). 

% -----------------------------------------------------

\subsection{Fusion cross sections for $^{132}$Sn+$^{48}$Ca and for $^{124}$Sn+$^{40}$Ca}

\begin{figure}[!htb]
\includegraphics*[width=8.6cm]{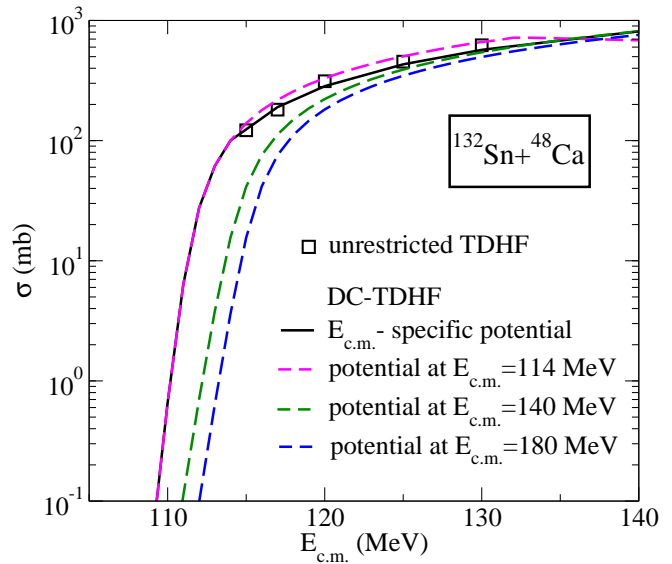}
\caption{\label{fig:sigma_fus_Sn132} (Color online) Total fusion cross section
for the neutron-rich system $^{132}$Sn+$^{48}$Ca. Compared are
DC-TDHF calculations using the energy-dependent heavy-ion potential
in Fig.~\ref{fig:pot1} with unrestricted TDHF calculations (see Fig.~\ref{fig:fus1})
at energies above the barrier.}
\end{figure}

In Fig.~\ref{fig:sigma_fus_Sn132} we show the total fusion
cross section for the neutron-rich system $^{132}$Sn+$^{48}$Ca.
The cross sections obtained with unrestricted TDHF runs above
the barrier (see Fig.~\ref{fig:fus1}) are indicated by squares. 
In addition, we display results obtained with the DC-TDHF method
which can be used at energies $E_{\mathrm{c.m.}}$ below and above
the potential barriers. The dashed lines correspond to calculations
using the transformed potential $U(\bar{R})$ in Fig.~\ref{fig:pot1}
at three different energies. We observe that at high energies,
$E_\mathrm{c.m.}=135-140$~MeV, all of the energy-dependent heavy-ion potentials
yield the same fusion cross section. However, these three potentials
yield substantially different cross sections at $E_\mathrm{c.m.}$ energies
near and below the barrier. In particular, we observe that the heavy-ion potential
calculated at the lowest possible energy $E_\mathrm{c.m.}=114$~MeV (with a barrier
height of $111.9$~MeV) gives by far the largest fusion cross section at
sub-barrier energies. According to theory, 
at a given $E_\mathrm{c.m.}$ energy, one should use the heavy-ion potential
calculated at this energy. We call this the ``$E_\mathrm{c.m.}$-specific potential''.
The corresponding total fusion cross sections are given by the solid black line
in Fig.~\ref{fig:sigma_fus_Sn132}, and we can see that these DC-TDHF results are
in perfect agreement with the unrestricted TDHF runs at above-barrier energies.

\begin{figure}[!htb]
\includegraphics*[width=8.6cm]{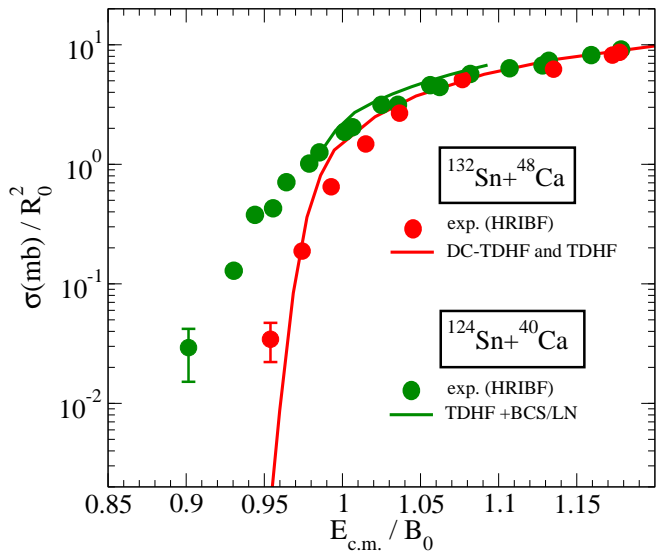}
\caption{\label{fig:sigma_scaled} (Color online) Comparison between scaled experimental
fusion cross sections~\cite{Li11a} and theoretical results obtained with the DC-TDHF
method and with unrestricted TDHF.}
\end{figure}

In Fig.~\ref{fig:sigma_scaled} we compare our theoretical fusion cross sections for
$^{132}$Sn+$^{48}$Ca and $^{124}$Sn+$^{40}$Ca to experimental
data obtained at HRIBF~\cite{Li11a,KR12}. The data for the latter system
agree with earlier measurements by Scarlassara {\it et al.}~\cite{Sc00}. In order to take out
trivial size-effects, the HRIBF experimentalists decided to scale their measured fusion
cross sections and center-of-mass energies as follows~\cite{Li11a}:
Defining the size parameter for the composite system
\begin{equation}
R_0 = A_1^{1/3} + A_2^{1/3}
\label{eq:R0}
\end{equation}
the fusion cross section is scaled according to
\begin{equation}
\sigma \rightarrow \sigma / R_0^2
\label{eq:sig_R0}
\end{equation}
Using the definition of the Coulomb barrier height parameter
\begin{equation}
B_0 = Z_1 Z_2 / R_0
\label{eq:B0}
\end{equation}
the center-of-mass energy is scaled as follows
\begin{equation}
E_{\mathrm{c.m.}} \rightarrow E_{\mathrm{c.m.}} / B_0
\label{eq:Ecm_B0}
\end{equation}
This scaling procedure yields the same total fusion cross section at
high energies for all measured Sn+Ca isotopes. Note that when one corrects
for nuclear size effects, the scaled fusion cross sections for the
neutron-rich system $^{132}$Sn+$^{48}$Ca are \emph{lower} than the fusion cross
sections for the stable $^{124}$Sn+$^{40}$Ca system, in contrast to the
absolute cross sections plotted in Fig.~\ref{fig:fus1}.
In the case of $^{124}$Sn+$^{40}$Ca, we are only able to compare
the experimental data to our unrestricted TDHF runs at above-barrier energies
(see Fig.~\ref{fig:fus1}). As we pointed out in Section III.A, these runs
were initialized with BCS/Lipkin-Nogami pairing for $^{124}$Sn, with the
BCS occupation numbers kept frozen during the collision. Unfortunately, this approximation
cannot be utilized in the DC-TDHF method because the static HF solution coupled with a constraint
on the instantaneous TDHF density for the combined system requires the reevaluation of the
occupation numbers for the lowest energy solution.
The good agreement between the measured fusion cross sections and our theoretical results
is quite remarkable because the only input in TDHF and DC-TDHF is the 
Skyrme N-N interaction with no adjustable parameters.
% ------------------------------------------------------------------------------

\section{Summary}

In this paper, we present dynamic microscopic calculations of 
pre-equilibrium giant dipole resonance excitation and fusion 
in the neutron-rich system $^{132}$Sn+$^{48}$Ca at energies near the
Coulomb barrier, and we compare photon yields and total fusion cross sections to those
of the stable system $^{124}$Sn+$^{40}$Ca. The numerical calculations
are carried out on a 3D lattice using both the Time-Dependent
Hartree-Fock (TDHF) method and the recently developed Density Constrained TDHF
method (DC-TDHF). 

Regarding pre-equilibrium GDR excitation, we demonstrate that the peak of the 
dipole radiation spectrum occurs at a substantially lower energy than expected
for an equilibrated system, thus reflecting the very large prolate elongation
of the di-nuclear complex during the early stages of fusion.

In addition, we study total fusion cross sections for both systems. First, we
carry out unrestricted TDHF runs at energies above the barrier (no tunneling).
Secondly, we calculate the heavy-ion interaction potential $V(R)$ and the
coordinate-dependent mass parameter $M(R)$ utilizing the DC-TDHF method.
Our results demonstrate that in these heavy
systems the potential barrier height increases dramatically with increasing
energy $E_\mathrm{c.m.}$, and the barrier peak moves inward towards
smaller $R$-values, in contrast to light ion systems such as $^{16}$O+$^{16}$O
which show almost no energy-dependence.

By numerical integration of the Schr\"odinger equation for the relative
coordinate $R$ with the {\it Incoming Wave Boundary Condition} (IWBC) method,
we calculate total fusion cross sections both below and above
the barrier. We emphasize that the only input in our microscopic calculations is the 
Skyrme N-N interaction, with no adjustable parameters whatsoever.
We find that our theoretical fusion cross sections for both
systems agree reasonably well with recent data measured at HRIBF.

% ------------------------------------------------------------------------------

\begin{acknowledgments}
This work has been supported by the U.S. Department of Energy under Grant No.
DE-FG02-96ER40963 with Vanderbilt University, and by the German BMBF
under Contracts No. 06ER9063  and No. 06FY9086. 
\end{acknowledgments}

% ------------------------------------------------------------------------------

\end{document}